\pdfoutput=1

\documentclass[11pt]{article}

\usepackage[T1]{fontenc}
\usepackage[utf8]{inputenc}
\usepackage{lmodern}
\usepackage[letterpaper,margin=1in]{geometry}
\usepackage{amsmath,amssymb,bm}
\usepackage{graphicx}
\usepackage{booktabs}
\usepackage{natbib}
\usepackage{microtype}
\usepackage{authblk}
\usepackage{url}
\usepackage{xurl}
\usepackage{caption}
\usepackage[hidelinks]{hyperref}
\usepackage{fancyhdr}

\title{\bfseries Hard conservation correctors can hide a degrading model when training autoregressive emulators}

\author[1,*]{William E. Chapman}
\author[2]{John S. Schreck}
\author[2]{Yingkai Sha}

\affil[1]{Department of Atmospheric and Oceanic Sciences, University of Colorado Boulder, Boulder, Colorado, USA}
\affil[2]{NSF National Center for Atmospheric Research, Boulder, Colorado, USA}
\affil[*]{Corresponding author: \href{mailto:wchapman@colorado.edu}{wchapman@colorado.edu}}

\date{}

\begin{document}

\maketitle

\begin{abstract}
AI weather and climate emulators increasingly incorporate physical principles into their formulation. One approach is to apply hard correctors that modify network outputs so that global mass, water, or energy budgets close. Prior work introduced such training-time correctors in the CREDIT framework and reported reduced precipitation bias and improved stability. Motivated by those results, we fine-tuned a global atmosphere emulator with a water-budget corrector, using the corrected prediction in the supervised loss and evaluating through post-correction budget closure. By that measure, training appeared successful. Every delivered field closed the moisture budget to machine precision. However, raw precipitation developed a growing global low bias over 18 training epochs, while the required correction increased from about 2\% to roughly 24\%. The cause is a scale degeneracy. A uniform change in raw precipitation amplitude is offset by a compensating change in the correction factor, leaving the corrected field, and therefore the supervised loss, unchanged. This invariance removes the restoring force on raw precipitation amplitude, allowing other training pressures to drive drift. Two changes recovered stable behavior. We supervised the pre-correction prediction and penalized its raw budget imbalance, while the hard correction remained in place for the delivered field. The required correction returned to less than 1\% within the next epoch. A controlled $2\times2$ ablation showed that the runaway occurred only when corrected-output supervision was combined with no imbalance penalty. Exact post-correction closure therefore says little about whether the raw model has learned the budget. When a corrector removes information from the loss, the raw fields and the applied correction need to be tracked.
\end{abstract}


\section{Introduction}\label{sec:intro}

AI weather and climate emulators are increasingly designed to conserve fundamental physical quantities. These constraints are often enforced using ``correctors" or ``fixers" that rescale or project network outputs to satisfy a prescribed physical law. Recent applications have used such correctors to close global mass, water, and energy budgets during training and inference, with reported benefits for the stability and skill of long-range simulations \citep{wattmeyer2025,sha2025,gregory2026floenet,chapman2025camulator}. However, exact closure of the corrected output does not necessarily ensure that the underlying network prediction remains physically meaningful.

Neural network emulators of weather and climate can match traditional numerical models on several deterministic forecast metrics \citep[e.g.,][]{bi2023,lam2023,kochkov2024} and increasingly target long, stable climate time-scale integrations \citep[e.g.,][]{bonev2023sfno, karlbauer2024healpix, wattmeyer2025, chapman2025camulator}, yet they do not inherit physical conservation laws by construction. A network can violate global mass, water, or energy budgets even when its pointwise error is small \citep{beucler2021,sha2025}. Soft constraints address this by adding the violation to the loss \citep{raissi2019physicsinformed,karniadakis2021physicsinformed,beucler2021}, whereas hard constraints project or rescale the output so that the constraint holds exactly \citep{beucler2021,zanetta2023,valente2025,harder2023hardconstrained,wattmeyer2025,sha2025}. The strategies can be combined, with a soft penalty shaping the learned prediction while a hard correction guarantees closure of the delivered field \citep{beucler2021}.

Hard correctors are attractive because they guarantee closure to machine precision without tuning a penalty weight. In the CREDIT framework, mass and energy budgets are closed by correctors that adjust surface pressure and atmospheric energy after the network forward pass \citep{schreck2025,sha2025, schreck2025controllable}, and a globally multiplicative precipitation corrector of the form introduced for the ACE models closes the water budget \citep{wattmeyer2023ace,wattmeyer2025}. Evaluating the supervised loss on the corrected output also appears reasonable. Every training sample satisfies the intended budget, and prior studies show that conservation imposed during training can improve stability and precipitation behavior \citep{wattmeyer2025,sha2025}.

Building on this work, we enabled a hard water-budget corrector while fine-tuning a global emulator designed to simulate atmospheric variability from days to centuries. Although every delivered field closed the moisture budget to machine precision, the pre-correction precipitation developed a large low bias that was invisible to the standard conservation diagnostic. We trace this failure to a scale degeneracy, or non-identifiable direction of the training objective, introduced by evaluating the supervised loss on the corrected output. Figure~\ref{fig:schematic} contrasts this configuration with training without a corrector and with an alternative formulation that combines pre-correction supervision, an imbalance penalty, and hard correction of the delivered field. We use this case to identify diagnostics that expose drift in the underlying network prediction even when the corrected output satisfies the intended constraint exactly.

\section{Data and Methods}\label{sec:methods}

\subsection{Emulator and training configuration}\label{sec:setup}

These experiments are produced on CAMulator, a global atmospheric emulator built on the WXFormer architecture of the CREDIT framework \citep{schreck2025} and trained to reproduce 6-hourly Community Atmosphere Model (CAM6) states on a $1.0^\circ$ grid with 32 hybrid sigma levels \citep[see][for full details]{chapman2025camulator}. Prognostic fields include winds, temperature, total specific humidity, surface pressure, and surface precipitation. Training minimizes a latitude- and variable-weighted error between the predicted and target next state. The experiments reported here come from a fine-tuning phase (CAMulator version~2) in which the decoder blocks (approximately 18\% of the model weights) that emit every prognostic and diagnostic field, precipitation included, were trainable while the backbone was frozen; optimization used Adam at a fixed learning rate of $2.5\times10^{-5}$ with no weight decay. Following \citet{sha2025} and \citet{chapman2025camulator}, this fine-tuned CAMulator is trained to respect global dry-mass, water, and energy budgets.

\subsection{Global water-budget corrector}

The water corrector computes the globally integrated tendency of total column water, evaporation source, and precipitation sink. It then rescales predicted precipitation by the factor required to close the budget. We denote this factor $r$, where $r=1$ means the raw prediction already closes. The fractional correction is $r-1$, reported below as a percentage. With evaporation and precipitation defined as positive source and sink magnitudes, respectively, the residual and sign convention are given in the appendix.

\subsection{Reference water budget}

The target data themselves require almost no correction. Applying the same calculation to one year of CAM6 output (1459 6-hourly steps in 1980, after pairing consecutive states for the tendency) gives a mean correction of 0.024\% with a standard deviation of 2.22\%. The global precipitation sink balances evaporation at $1.72\times10^{10}$~kg~s$^{-1}$ (annual average) with near-zero mean storage. Persistent model corrections of order 10\% therefore do not originate in the target data. The 2.22\% per-step spread reflects the 6-hourly temporal discretization of the budget from time-averaged output, not a CAM6 conservation error.

\subsection{Training interventions and ablation design}\label{sec:fix}

With the reference budget established, we now describe the training configurations that produced the failure and its remedy. The baseline production configuration retained the hard water-budget corrector, evaluated supervised precipitation loss on the corrected output, and applied no raw-imbalance penalty (Fig.~\ref{fig:schematic}b). This is the same experimental setup used in \citet{sha2025} and \citet{chapman2025camulator}.

After the required correction had grown to approximately 24\%, we resumed training from that checkpoint with two changes (Fig.~\ref{fig:drift}a). First, the supervised loss was moved from the corrected output to the pre-correction prediction, restoring direct supervision of raw precipitation amplitude. Second, we added a soft penalty on the squared raw closure deviation, $(r-1)^2$, to encourage the network to reduce its own water-budget imbalance. The hard corrector remained active in the forward pass so that the delivered field continued to close the budget exactly (Fig.~\ref{fig:schematic}c). In essence, this moves the configuration, at epoch 18, from Fig.~\ref{fig:schematic}b to Fig.~\ref{fig:schematic}c.

Because the final network configuration changed both supervised-loss placement and the penalty simultaneously, we performed a controlled $2\times2$ ablation to separate their effects (Fig.~\ref{fig:ablation}). Four short fine-tuning training cycles, from a common checkpoint, crossed the supervised target (corrected versus pre-correction prediction) with the penalty weight (0 versus 0.1), giving cells~A (corrected output, no penalty), B (corrected output, penalty), C (pre-correction, no penalty), and D (pre-correction, penalty). The ablation used a configuration that drifted faster than the production run, so its correction magnitudes are not directly comparable to the production trajectory in Fig.~\ref{fig:drift}a. We did not test inference-only correction because prior work found training-time conservation important for forecast stability \citep{sha2025}.

\par\medskip
\noindent\begin{minipage}{\linewidth}
\centering
\includegraphics[width=\linewidth]{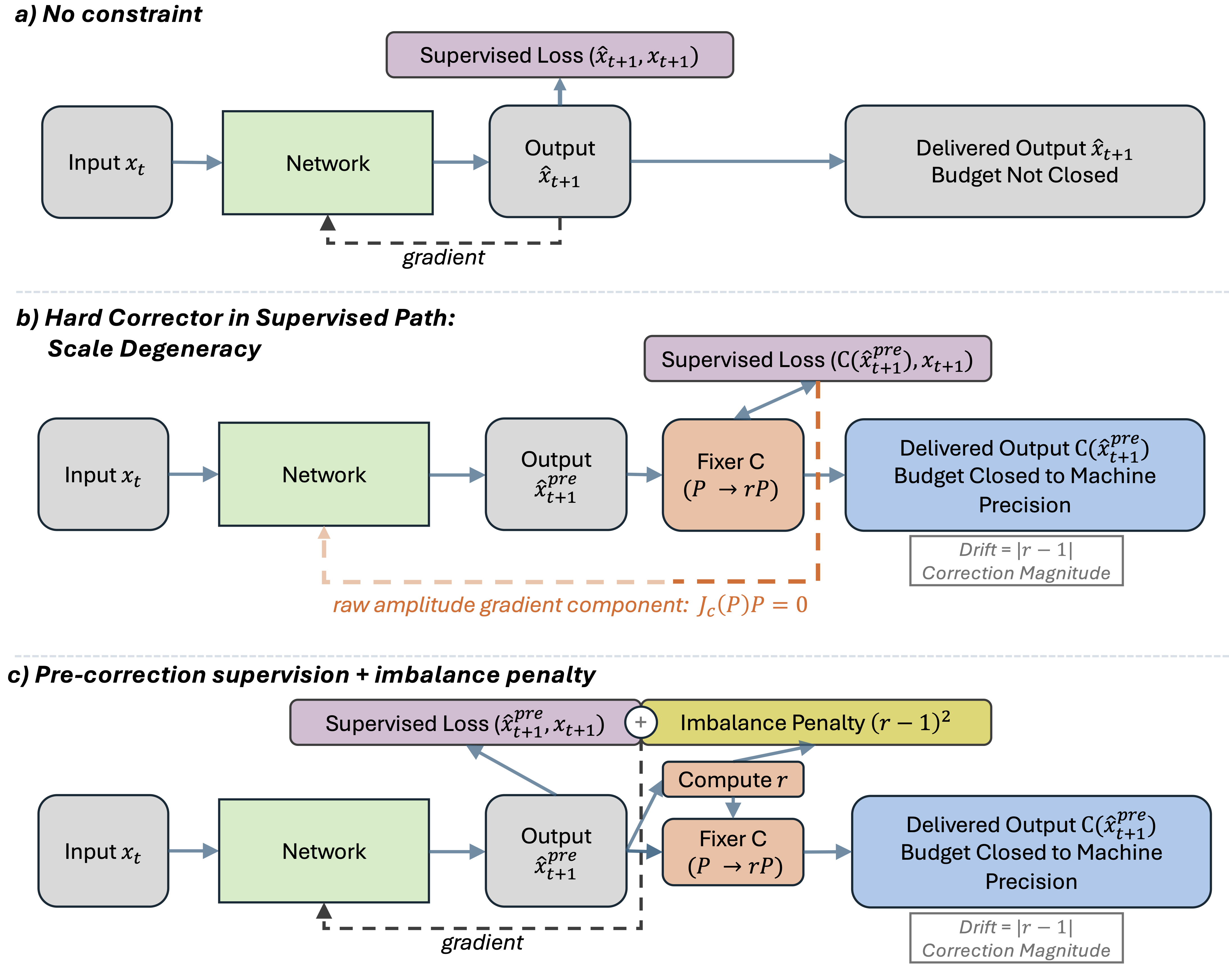}
\captionof{figure}{Schematic of the three water-budget training configurations. (a)~The raw network prediction is supervised directly, with no budget correction. (b)~The water-budget corrector is applied before the supervised loss, so training is based on the corrected prediction. (c)~The raw prediction is supervised and penalized for budget imbalance, while the corrector is retained for the delivered output. In both corrected configurations, precipitation is rescaled by the budget-derived factor $r$ so that the delivered field closes the global water budget exactly.}
\label{fig:schematic}
\end{minipage}
\medskip

\section{Results}\label{sec:results}

\subsection{The required correction and the raw precipitation bias}\label{sec:growth}

We first examine the baseline production configuration, which used corrected-output supervision with no penalty (Fig.~\ref{fig:schematic}b). During fine-tuning the required correction grew from 2.4\% to 24.1\% over 18 epochs (Fig.~\ref{fig:drift}a, orange), while the delivered field continued to close the global water budget to machine precision at every step. The failure was easy to miss because the corrector rescaled predicted precipitation before the supervised error was evaluated, so the standard conservation diagnostic reported success throughout, and the deterioration surfaced only in quantities outside that pass-fail test. Over the same span the raw global precipitation error worsened from $-4.0\%$ to $-19.8\%$ of the CAM6 target, whereas the corrected error, the delivered post-correction precipitation measured against the same target, changed only from $-1.7\%$ to $-0.5\%$ because the corrector sets that delivered total from the predicted evaporation and storage; the raw precipitation sink itself declined steadily toward a low plateau (Fig.~\ref{fig:drift}c). The model was still delivering an exactly conservative field, but only because the corrector was compensating for an increasingly biased raw prediction.

Four checks ruled out alternative sources of the drift. The same budget calculation applied to CAM6 showed near-zero mean imbalance, so the trajectory did not originate in the target data. A second training run in the same configuration reproduced the behavior and drifted further, with per-step corrections peaking above 60\%, ruling out a single anomalous optimization path. The mass and energy correctors remained near 0.01\% while the water correction grew (Fig.~\ref{fig:drift}d), indicating that the model had not simply become unstable in every conserved quantity. Finally, recomputing the global reductions at higher precision changed the correction by only tenths of a percent, far too little to explain a 24\% signal (see appendix for details). Together, these checks pointed to the coupling between the precipitation corrector and the supervised loss.

\par\medskip
\noindent\begin{minipage}{\linewidth}
\centering
\includegraphics[width=0.96\linewidth]{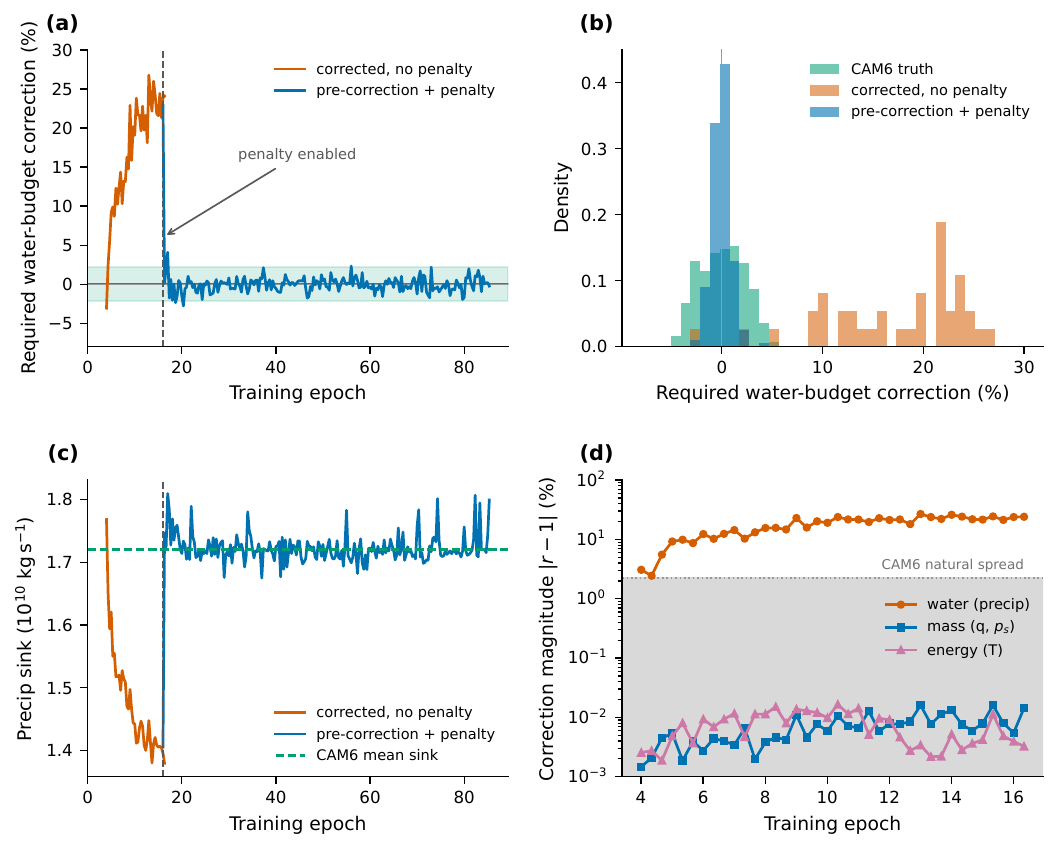}
\captionof{figure}{Evolution of the water-budget correction during fine-tuning. (a)~The required correction grows to about 24\% when the supervised loss is evaluated on the corrected output, then returns to the CAM6 reference range after switching to pre-correction supervision with an imbalance penalty. The dashed line marks this change in training configuration. (b)~Distributions of the required correction for CAM6 and the two training configurations. (c)~The corresponding raw global precipitation sink, with the CAM6 annual mean shown by the dashed line. (d)~Mass, water, and energy correction magnitudes during the corrected-output run. Only the water correction exhibits sustained growth.}
\label{fig:drift}
\end{minipage}
\medskip

\subsection{Scale degeneracy}\label{sec:mechanism}

A loss evaluated only on the corrected field carries no information about the raw global precipitation amplitude, because any uniform rescaling produces the same corrected output. The corrector rescales raw precipitation to whatever global total the budget demands. If the network doubled its entire precipitation field, the correction factor would halve and the delivered output would be unchanged.

Formally, let $P$ be the raw precipitation field, let $\mathbf{1}^{T}P$ denote its global total, with grid-cell area weights absorbed into the notation, and let $Q$ be the total precipitation required by the predicted storage tendency and evaporation. The correction operator $C$, shown as ``Fixer~$C$'' in Fig.~\ref{fig:schematic} and defined in full in the appendix, preserves the normalized spatial pattern $P/(\mathbf{1}^{T}P)$ while replacing its global amplitude with $Q$:
\begin{equation}
C(P)
=
Q\frac{P}{\mathbf{1}^{T}P},
\qquad
C(\alpha P)
=
C(P),
\qquad
J_C(P)P
=
0,
\label{eq:null}
\end{equation}
for any positive scalar $\alpha$. The final equality follows because a perturbation along $P$ gives $P+\epsilon P=(1+\epsilon)P$, which is only a uniform rescaling and therefore leaves the corrected field unchanged. Thus, the raw-amplitude direction lies in the null space of the correction Jacobian, and a supervised loss evaluated only on $C(P)$ provides no restoring gradient along that direction.

We verified the invariance by uniformly rescaling CAM6 precipitation with $\alpha=0.5$, 0.75, 1.0, 1.25, and 1.5. The corrected field $C(\alpha P)$ was unchanged to machine precision, the correction factor scaled as $r\propto1/\alpha$, and the corrected-output loss remained flat, whereas the raw-output loss and imbalance penalty responded strongly. A finite-difference probe along $P$ likewise remained at roundoff, consistent with $J_C(P)P=0$. This test isolates precipitation scaling alone; during training, precipitation, evaporation, and storage change jointly.

The scale invariance explains why the corrected loss cannot restore raw amplitude, but not why the observed drift was consistently downward. We did not isolate that forcing. Plausible contributors include gradients from strongly supervised fields that share decoder weights with precipitation and the tendency of mean-squared-error training to favor under-prediction for displaced, heavy-tailed precipitation \citep{gilleland2009,subich2025}. Weight decay was disabled. The apparent plateau likely marks departure from uniform rescaling, which reintroduces opposing gradients through changes in the spatial pattern or other jointly predicted fields.

The contrast among correctors helps explain why the vulnerability became visible only for water. Mass and energy adjustments act on temperature, humidity, and surface pressure, which remain supervised across many levels or through coupled terms. After the water adjustment rescales precipitation, no comparable supervision remains on its global amplitude. Because the operators also have different null spaces, weak supervision alone is not a complete explanation. A null direction produces visible runaway only when it overlaps a degree of freedom that the remaining objective does not strongly constrain.

\subsection{Pre-correction supervision with a penalty}\label{sec:recovery}

Starting from the degenerate checkpoint, the stable configuration described in Section~\ref{sec:fix}, with penalty weight 0.1, reduced the required correction from 24\% to below 1\% within one epoch and held it near zero for the next 64 epochs; over the final five epochs the mean correction was 0.4\% with a standard deviation of 0.9\% (Fig.~\ref{fig:drift}b). Because the soft penalty targets $(r-1)^2$ directly, the collapse of the required correction is necessary but not sufficient evidence of an improved model. A partly independent indicator is the recovery of the raw precipitation sink toward the CAM6 reference (Fig.~\ref{fig:drift}c), although this indicator is independent only to the extent that the supervised evaporation and storage tendency are unbiased. We did not measure raw-field forecast skill.

\subsection{Ablation study}\label{sec:ablation}

The controlled ablation isolated which change prevented the runaway and exposed its signature in the raw field. It crosses two factors, the supervised target (corrected output versus the pre-correction prediction) and the imbalance-penalty weight (0 versus 0.1), giving the four cells A--D defined in Section~\ref{sec:fix}; cell~A is the production failure mode, corrected-output supervision with no penalty. Only cell~A drifted, its required correction rising to 60--70\% within an epoch and then partially receding (Fig.~\ref{fig:ablation}a), a third independent occurrence of the runaway alongside the two production runs. The same cell shows the failure directly in the raw prediction, its global precipitation sink collapsing to roughly $1.0\times10^{10}$~kg~s$^{-1}$, some 40\% below the CAM6 reference, while the three stable cells hold near it (Fig.~\ref{fig:ablation}b).

The other three cells stayed centered near zero across four epochs with no systematic growth, at per-step means of 1.1\%, 0.5\%, and 0.1\% for cell~C (pre-correction, no penalty), cell~B (corrected output, penalty), and cell~D (pre-correction with penalty, the production remedy), respectively. These are single runs, and the reported spreads (1.6--3.1\%) are within-run step-to-step variation rather than across-seed uncertainty, so the three stable cells are not distinguishable from one another at this sample size; the resolved contrast is the runaway cell against the other three. The runaway therefore requires the specific pairing of corrected-output supervision with no competing constraint, and restoring pre-correction supervision or adding the penalty each suppresses it on its own.

\par\medskip
\noindent\begin{minipage}{\linewidth}
\centering
\includegraphics[width=0.96\linewidth]{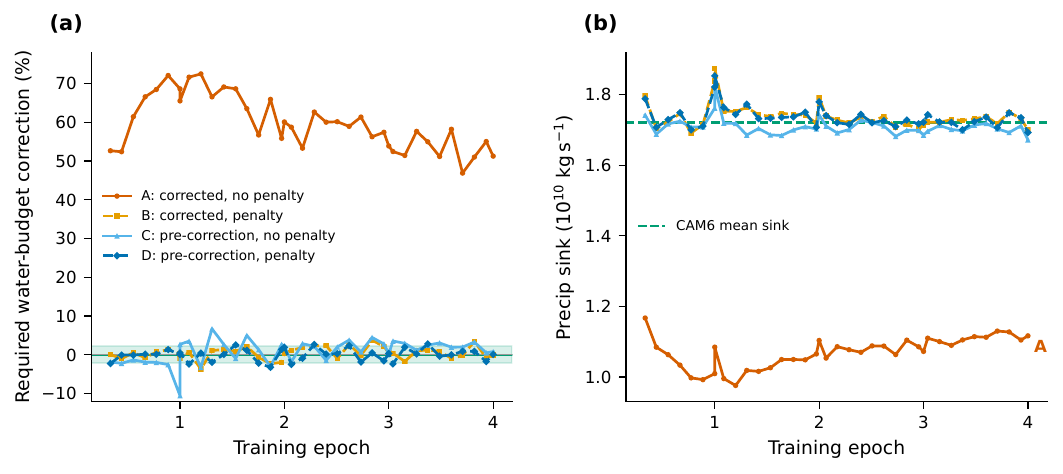}
\captionof{figure}{Results of the $2\times2$ ablation crossing the supervised target, corrected or pre-correction, with an imbalance-penalty weight of 0 or 0.1. (a)~The required water-budget correction grows only for corrected-output supervision without a penalty (cell A); the other three configurations remain near the CAM6 reference range. (b)~The raw global precipitation sink for the same experiments. Cell A develops a large low bias, while the other configurations remain near the CAM6 mean. Shading and error bars show within-run step-to-step variability.}
\label{fig:ablation}
\end{minipage}
\medskip

\section{Discussion and Conclusions}\label{sec:discussion}

Hard constraint correctors can be used successfully within differentiable architectures \citep{beucler2021,zanetta2023,valente2025,harder2023hardconstrained}. The specific danger occurs when the corrected field is the sole supervised target and the correction operator has a nontrivial null direction corresponding to a physically meaningful property of the raw prediction. The loss then cannot identify that property unless another term constrains it.

This yields a general diagnostic beyond multiplicative precipitation correction. Let $C(\hat x)$ be a differentiable correction and let the supervised objective be $\mathcal{L}_{\mathrm{sup}}(x_{t+1},C(\hat x))$. Any perturbation $v$ satisfying $J_C(\hat x)v=0$ is locally invisible to that objective because
\begin{equation}
\left.\frac{d}{d\epsilon}\mathcal{L}_{\mathrm{sup}}\!\left(x_{t+1},C(\hat x+\epsilon v)\right)\right|_{\epsilon=0}
=\nabla_C\mathcal{L}_{\mathrm{sup}}^{T}J_C(\hat x)v=0.
\label{eq:generalnull}
\end{equation}
If $v$ represents a physically meaningful property and no other term identifies it, the raw model is nonidentifiable in that direction. Before training, the key question is which degrees of freedom the correction hides from the loss.

For a multiplicative global rescaling, Eq.~\eqref{eq:null} shows that raw amplitude is invisible after correction, so during training that amplitude must be constrained independently. The configuration of Section~\ref{sec:fix} does this by supervising the pre-correction prediction. Corrected-output supervision combined with a penalty also controls the degeneracy in our ablation (Section~\ref{sec:ablation}), but that route stays tuning dependent through the penalty weight, whereas pre-correction supervision constrains amplitude directly.

The result applies most directly to deterministic rescalings and projections with null spaces that overlap weakly supervised model degrees of freedom, including global moisture, energy, mass, and surface-pressure fixers. It should not be generalized automatically to every constraint operator.

Evaluation must follow the same logic. After a hard correction, a closed budget is no longer an independent model diagnostic. We therefore track the raw residual and the size of the correction the model requires, alongside skill in both the raw and corrected fields. A growing correction can reveal deterioration even while the delivered field remains exactly conservative.

The raw drift still matters even when the corrected global mean remains accurate. A large correction means that precipitation amplitude is increasingly set by predicted evaporation and storage rather than by the raw precipitation field itself. The corrector also adjusts only one global scalar, so it cannot repair regional errors once the drift departs from uniform rescaling. In free-running simulations, the corrected state feeds back at every step, and stability under large repeated corrections remains untested. We have not shown that the corrected product degrades. The cost is the lost diagnostic information, and the risk that drift carries the model outside the correction null space.

Our result complements \citet{sha2025}, who report benefits from global conservation during training and inference, including reduced precipitation bias and improved long-range stability. Those results show that training-time conservation can be beneficial. Our case shows one failure mode that appears when the corrected output is the only supervised target.

The Ai2 Climate Emulator (ACE2) provides a close comparison \citep{wattmeyer2025}. Its corrector multiplies precipitation by a globally uniform factor to close the moisture budget (the same operator we analyze) and is applied before a mean-squared-error loss with no conservation penalty. ACE2 reports stable long integrations, with no drift in total atmospheric moisture and a realistic precipitation climatology. These diagnostics are computed after correction, however, and do not include the correction factor or the raw precipitation field. ACE2 therefore contains the same precipitation-amplitude null direction, but its published diagnostics do not establish whether training produced meaningful drift along that direction, which is exactly the post-correction blind spot we document. The comparison is at the level of diagnostics, since our measured drift is a per-training-step quantity whereas ACE2's reported stability is free-running and climatological. We do not claim that ACE2 exhibits the raw-amplitude degradation documented here.

Our evidence is limited to one emulator and one water-budget corrector. The stable production run changed supervised-loss placement and penalty weight at once. The controlled $2\times2$ in Section~\ref{sec:ablation} separates them and shows that either change alone prevents the runaway, but those four runs span four epochs from a single checkpoint, so they establish the short-term mechanism rather than long-term climate skill.

The reported correction is measured per training step, and we do not claim improved forecast skill or climate fidelity. The stable configuration remained near zero for 64 epochs, but longer free-running integrations are required to assess accumulated drift. The CAM6 reference is stationary across 1980--84, with annual mean corrections from 0.022\% to 0.029\% and standard deviations from 2.04\% to 2.29\%. Any broader skill benefit rests on prior work rather than the experiments reported here, since we did not compare inference-only correction against training-time penalization.

\section*{Acknowledgments}
We acknowledge high-performance computing support from the Casper and Derecho systems provided by the Computational and Information Systems Laboratory (CISL) at the NSF National Center for Atmospheric Research (NCAR). Derecho is identified by DOI \url{https://doi.org/10.5065/qx9a-pg09}. This paper is under review for a Lessons Learned manuscript at Artificial Intelligence for the Earth Systems an AMS Journal. The authors used OpenAI's and Anthropic's language model (ChatGPT, \url{https://chat.openai.com/} and Claude \url{https://claude.ai/}, last access: 8 July 2026) to assist with grammar, phrasing, and consistency checks during manuscript preparation and Claude Code for figure paneling.

\section*{Data and code availability}
The model code, configurations, job scripts, and figure-generation code are available at \url{https://github.com/WillyChap/miles-credit/tree/camulator-conservation-lessons}, with the experiments organized under \path{experiments/}; the figures rebuild from cached data in \path{experiments/figure_data/}. The CAM6 training data are Community Earth System Model output.

\appendix
\section{The conservation-coupled training loss}\label{app:loss}

Let $x_t$ be the input state, $\hat x_{t+1}$ the network prediction of the next state, and $x_{t+1}$ the target next state. The supervised term is a latitude- and variable-weighted error,
\begin{equation}
\mathcal{L}_{\mathrm{sup}}(x_{t+1},\hat x_{t+1})=\frac{1}{V}\sum_{v=1}^{V}\frac{1}{BHW}\sum_{b,i,j}w_vL_i^{\mathrm{lat}}\ell\!\left(x_{b,v,ij},\hat x_{b,v,ij}\right),
\end{equation}
where $L_i^{\mathrm{lat}}=\cos(\varphi_i)^p/\overline{\cos(\varphi)^p}$ and $\ell$ is mean squared error.

For each sample $b$, the water corrector forms globally integrated total-column-water tendency, positive evaporation source, and positive precipitation sink,
\begin{equation}
\mathcal{T}_b=\!\sum_{ij}a_{ij}\frac{\mathrm{TWC}^{\mathrm{pred}}_{b,ij}-\mathrm{TWC}^{\mathrm{in}}_{b,ij}}{\Delta t},\quad
\mathcal{E}_b=\!\sum_{ij}a_{ij}\frac{E_{b,ij}\rho_w}{\Delta t},\quad
\mathcal{P}_b=\!\sum_{ij}a_{ij}\frac{P_{b,ij}\rho_w}{\Delta t}.
\end{equation}
Closure requires $\mathcal{T}_b=\mathcal{E}_b-\mathcal{P}_b$. The residual and precipitation correction factor are therefore
\begin{equation}
R_b=\mathcal{E}_b-\mathcal{T}_b-\mathcal{P}_b,\qquad
r_b=\frac{\mathcal{P}_b+R_b}{\mathcal{P}_b}=\frac{\mathcal{E}_b-\mathcal{T}_b}{\mathcal{P}_b},\qquad
r_b-1=\frac{R_b}{\mathcal{P}_b}.
\end{equation}
The corrector applies $P\!\to\!r_bP$. The soft penalty is the normalized raw imbalance,
\begin{equation}
\mathcal{L}_{\mathrm{water}}=\frac{1}{B}\sum_{b=1}^{B}(r_b-1)^2=\frac{1}{B}\sum_{b=1}^{B}\left(\frac{R_b}{\mathcal{P}_b}\right)^2.
\end{equation}
The stable objective is
\begin{equation}
\mathcal{L}=\mathcal{L}_{\mathrm{sup}}\!\left(x_{t+1},\hat x^{\mathrm{pre}}_{t+1}\right)+\lambda_w\mathcal{L}_{\mathrm{water}},
\label{eq:total}
\end{equation}
where $\hat x^{\mathrm{pre}}_{t+1}$ is the pre-correction prediction and $\lambda_w=0.1$ in the stable runs reported here. Evaluating the supervised term only on corrected precipitation removes the uniform-amplitude direction from that term, as shown by Eq.~\eqref{eq:null}; it does not by itself choose the direction of drift. Equation~\eqref{eq:total} restores direct supervision of raw amplitude and adds a gradient toward raw budget closure. Mass and energy correctors compute analogous factors but contribute no penalty here.

\subsection{Implementation and numerical precision}

Two implementation details affect the correction diagnostic. Accumulating global reductions at bfloat16 precision corrupts the diagnostic. Across 30 sampled steps, bfloat16 introduced correction errors averaging 0.22\% and reaching 0.52\%. This is a monitoring hazard, because the errors are roughly fifty to a hundred times smaller than the drift documented in Section~\ref{sec:growth}. Float32 was adequate in our tests, and we used float64 to make numerical error negligible. The corrector must also return the raw imbalance explicitly so that it can be monitored and included in the loss.

\bibliographystyle{plainnat}
\bibliography{references}

\end{document}